# System Approach to Synthesis, Modeling and Control of Complex Dynamical Systems

ARMEN G. BAGDASARYAN
Institution of the Russian Academy of Sciences
V. A. Trapeznikov Institute for Control Sciences of RAS
65 Profsoyuznaya, 117997 Moscow
RUSSIA
abagdasari@hotmail.com

*Abstract:* - The basic features of complex dynamical and control systems, including systems having hierarchical structure, are considered. Special attention is paid to the problems of synthesis of dynamical models of complex systems, construction of efficient control models, and to the development of simulation techniques. An approach to the synthesis of dynamic models of complex systems that integrates expert knowledge with the process of modeling is proposed. A set-theoretic model of complex system is defined and briefly analyzed. A mathematical model of complex dynamical system with control, based on aggregate description, is also proposed. The structure of the model is described, and architecture of computer simulation system is presented, requirements to and components of computer simulation systems are analyzed.

## 1 Introduction

The science of complex systems is a multidisciplinary field aiming at understanding the complex real world that surrounds us. Examples of these systems are neural networks in the brain that produce intelligence and consciousness, artificial intelligence systems, swarm of software agents, social insect (animal) colonies, ecological systems, traffic patterns, biological systems, social and economic systems and many other scientific areas can be considered to fall into the realm of complex systems.

Complex systems contain a large number of mutually interacting entities (components, agents, processes, etc.) whose aggregate activity is nonlinear, not derivable from the summations of the activity of individual entities, and typically exhibit hierarchical self-organization. Another important characteristic of complex systems is that they are in some sense purposive. The description of complex systems requires the notion of purpose, since the systems are generally purposive. This means that the dynamics of the system has a definable objective or function [1].

Each element of a complex system interacts with other elements, directly or indirectly. The actions of or changes in one element affect other elements. This makes the overall behavior of the system very hard to deduce from and/or to track in terms of the behaviour of its parts. This occurs when there are many parts, and/or when there many interactions between the parts. Since the behavior of the system depends on the elements interactions, an integrative system-theoretic (top-down) approach seems more promising, as opposed to a reductionist (bottom-up) one.

Any scientific method (approach) of studying complex real world systems relies on modeling (analytical, numerical) and computer simulation.

The study of complex systems begins from a set of models that capture aspects of the dynamics of simple or complex systems. These models should be sufficiently general to encompass a wide range of capabilities but have sufficient structure to capture interesting features.

Most of the complex systems can be studied by using nonlinear mathematical models, statistical methods and computer modeling approaches. For this both analytical tools and computer simulation are adopted. Among the analytical techniques are statistical mechanics, stochastic dynamics, non-equilibrium thermodynamics, etc [29], [31], [36]. Among the computer simulation techniques are cellular automata, multi-agent techniques, evolutionary programming, Monte Carlo methods, etc, [3], [4], [5], [25]. However, the analytical methods alone do not allow us to understand a complex system. Since analytical treatments do not yield complete theories of complex systems, computer simulations play a key role in our understanding of how these systems function and work. This is also true and possibly in a more degree for complex control systems.

The main characteristic of modern complex control systems is that it is impossible to uniquely and adequately describe these systems, using classical mathematical methods. Classical mathematical models and approaches are yet suitable and applicable just for a few problem domains, which are static and comprehensible, and have most general properties. But in complex dynamic environments, with an increase of complexity, problem domains become dynamic, requiring for dynamic solutions that will be able to adapt to the changes in the problem domain, and there still remains a wide range of problems that can not be described by the existing formal methods.

## 2 The Properties of Complex Systems and Synthesis of Dynamic Models

Basic reasons that make it difficult for complex (control) systems to be described by formalized methods are the following ones:
- Information *incompleteness* on the state and the behavior of a complex system;
- Presence of a *human* (observer) as an intelligent subsystem that forms requirements and makes decisions in complex systems;
- *Uncertainty* (inconsistency, antagonism) and multiplicity of the purposes of a complex system, which are absent in a precise formulation;
- *Restrictions* imposed on the purposes (controls, behavior, final results) externally and/or internally in relation to a system are often unknown;
- *Weak structuredness*, uniqueness, combination of individual behaviors with collective ones are the intrinsic features of complex systems.

Complex systems are different from simple systems by their capabilities of:
- *Self-organization* – the ability of a complex system to autonomously change own behavior and structure in response to events and to environmental changes that affect the behavior.

For systems with a network structure, including hierarchical one, self-organization can amount to: (1) disconnecting certain constituent nodes from the system, (2) connecting previously disconnected nodes to the same or to other nodes, (3) acquiring new nodes, (4) discarding existing nodes, (5) acquiring new links, (6) discarding existing links, (7) removing or modifying existing links.
- *Co-evolution* - the ability of a complex system to autonomously change its behavior and structure in response to changes in the system environment and in turn to cause changes in the environment by its new (corrected) behavior.

Complex systems co-evolve with their environments: they are affected by the environment and they affect their environment.
- *Emergence* – the property that emerge from the interaction of constituent components of a complex system.

The emergent properties do not exist in the components and because they emerge from the unpredictable interaction of components they can not be planned or designed.
- *Adaptation* – the ability of a complex system to autonomously adjust its behavior in response to the occurrence of events that affect its operation.

Complex systems should adapt quickly to unforeseen changes and/or unexpected events in the environment. Adaptation enables the system to modify itself and to revive in changing environment.
- *Anticipation* – the ability of a system to predict changes in the environment to cope with them, and adjust accordingly.

Anticipation prepares the system for changes before these occur and helps the system to adapt without it being perturbed.
- *Robustness* - the ability of a system to continue its functions in the face of perturbations.

Robustness allows the system to withstand perturbations and to keep its function and/or follow purposes, giving the system the possibility to adapt.

Being oriented on the analysis of complex object as a whole, the system approach does include the methods of decomposition of complex system on separate subsystems. But the main purpose is the subsequent synthesis of subsystems, which provides the priority of a whole. However, to reach this priority is not simple. For a number of complex systems, optimum of the whole system can not be obtained from optimums of its subsystems. It should be noted, that complex systems that possess the property of integrity do not have constituent elements and act as one whole object. In this kind of systems, the connections and relations are so complicated and strong (all-to-all) that they can not be considered as an interaction between the localized parts of system. In physical systems the integrity corresponds to locality that is to such an influence of one part of system to another that can not be explained by interaction between them. As a rule, connections in integrative systems are often based on structural principles, but not on the cause/effect principle.

One of the approaches to synthesis of dynamic models of complex systems can be based on the use of the so-called *master-systems* consisting of canonical templates and expert knowledge-bases. The knowledge-base consisting of declarative and procedural knowledge realizes a conceptual model of complex system. The declarative knowledge should contain:
- Objectives tree of complex system that provides a decomposition of global goal on subgoals and description of relation between them;
- The architecture and/or structure of complex system;
- The set of canonical templates;
- The set of models of canonical templates;
- Problem domain databases.

The information about objectives tree can be represented in the form
$O = (I, Id, G, L)$, where
$I$ is a structure of global goal decomposition,
$Id$ is a structural identifier of nodes,
$G$ is a goal assigned to the node,
$L$ is a some rule/law describing the connection between the neighbor nodes.

The canonical template can be realized with the use of the language of system dynamics [6] or system state diagrams [18]. The canonical template has a certain structure, a set of input, output, and initial values/conditions. Formally the canonical template can be described as
$C = \{Str, T, X, Y, Iv, Tr\}$, where
$Str$ - a structure of the template,
$T$ - a rule/law of template functioning/dynamics,
$X$ - a set input parameters,
$Y$ - a set of output parameters,
$Iv$ - a set of initial values/conditions,
$Tr$ - rules of transformation of template structure, adding/modifying/removing links and/or nodes.

The canonical template is a separate object having information about its components and certain internal structure. The canonical template model is the object containing information not only about its components and structure but also the concrete input and output values and values of initial conditions.

Each canonical model is assigned to one of the goals of the objectives tree.

The synthesis of dynamic models of complex system is realized by transformation of declarative knowledge about problem domain to the algorithms of system state dynamics with the help of procedural knowledge.

The procedural knowledge is realized in knowledge-bases in the form of inference rules. They formalize the process of dynamic models synthesis. The inference rules provide the mapping of structure of conceptual model into the structure of dynamic models. The knowledge-base can contain different groups of inference procedures, depending on the purposes of investigation. For example, they can be of the following types: correspondence rules that determine for each canonical model the goal problems it solves; the inference rules that define informational relations between the templates in canonical model, etc.

The representation of conceptual model of complex system in the form knowledge-bases provides the autonomous usage of expert knowledge upon synthesizing the dynamic models.

The above models can be extended by adding to the canonical templates a set of input control symbol, thus providing the dynamic models with the mechanism of system control.

It is well-known that to control is in some sense to anticipate. For modeling and analysis of complex systems in the presence of principally non-formalizable problems and impossibility of strict mathematical formulation of a problem, heuristic, cognitive and robust approaches and methods can be applied.

However, if there exists a sufficiently exact and adequate formal description of a complex system (object, process) and we have a sufficient initial information, then heuristics, cognition and robustness is not required. But these are of a great need when there is a lot of uncertainty in the description of a complex system (object, process); a system has conflicting and/or inconsistent purposes; lack of initial information; problems are ill-posed.

## 3 Complexity, Modeling and Control

Complex systems are more often understood as dynamical systems with complex, unpredictable behavior. Multidimensional systems, nonlinear systems or systems with chaotic behavior, and also the systems, which dynamics depends on or determined by human being(s), are the formal examples of complex systems. In connection with

modeling and control complexity, complex systems have specific characteristics, among them:
- uniqueness;
- weak structuredness of knowledge about system;
- the composite nature of system;
- heterogeneity of elements composing the system;
- the ambiguity of factors affecting the system;
- multivariation of system behavior;
- multicriteria nature of estimations of system's properties; and, as a rule,
- high dimensionality of system.

Under such conditions, the key problem of complex systems theory consists in the development of methods of qualitative analysis of the dynamics of such systems and in the construction of efficient control techniques. In a general case, the purpose of control is to bring the system to a point of its phase space which corresponds to maximal or minimal value of the chosen efficiency criterion. Another one of the main and actual problems in the theory of complex systems and control sciences is a solution of "ill-posed, weakly- and poorly-structured and weakly-formalizable complex problems" associated with complex technical, organizational, social, economic, and cognitive and many other objects, and with the perspectives of their evolution. Since the analysis and efficient control are impossible without a formal model of a system, the technologies for construction (building) of models of complex systems have to be used.

Complexity of a system is a property stipulated by an internal law of the system that defines some important parameters, including spatial structure and properties of the processes in this structure. This definition of complexity is understood as certain physical characteristic of nature.

Since it is nonlinearity of internal regularities (laws) that underlies the complexity of real world systems, complexity and nonlinearity are sometimes considered as synonyms. The more complex a process or geometrical form of a system (object), the more it is nonlinear.

Today we can distinguish several basic forms of complexity: structural (geometrical, topological), dynamical, hierarchical, and algorithmic. However, other possible forms of complexity can be found as well. For example, one that comes from large scales. Algorithmic complexity finds itself in many software systems. These are the most complex systems developed by human being, although their structure and dynamics are comparatively simple. Structural, dynamical, algorithmic, hierarchical and large-scale complexity of systems attracts much of attention because we face them, manifestations of nonlinearity of nature, in our everyday life.

Interplay between intellectualized mathematical and information technologies of control and decision support plays an important role in modeling of processes of evolution and functioning of complex systems. Intellectualization of complex control systems is being actively developed in the last decade. In order to intellectualize modern control systems the artificial intelligence methods or intelligent subsystems embedded in control system are more often applied. The intellectualization of complex systems seems to be a positive and very perspective direction of control systems development. It significantly eases control decision making, as the underlying mechanisms are similar to those used by human intellect. Control processes in intellectualized systems are based on the experience, skills and knowledge, that is on the "understanding" of complex situations of purposeful behavior. An elementary intellect in control systems is constructed by using feedback loops and information flows, which give a system the capability of "understanding" of current situations. To "understand" the more complex situations, an adaptive subsystem should be added to the main feedback loop. However, if the object/system performs multiple functions (multipurpose object) then each function makes the system more complex and, as a consequence, the more sophisticated intelligent subsystems have to be used. But many questions remain: how to differ control systems by the level of intelligence; at what level of evolution of control systems they can be considered as artificial intelligence systems; what is the relation between adaptive properties and intelligent systems?

Complex systems are usually difficult to model, design, and control. There are several particular methods for coping with complexity and building complex systems.

At the beginning, a conceptual model of system is developed, which reflects the most important, in the context of the problem under study, material and energy and information processes taking place between different elements of system (that is, its subsystems), internal states of which can be considered as independent. This kind of model determines the general structure of system and it should be complemented by algorithmic and, more often, by mathematical models of each of subsystems. These models can be represented by graph models, Petri nets models, system dynamics models or by their combination. The obtained models are aggregate that reflect the dynamics of the most important, for current investigation, variables. Then, the next step consists in checking the mathematical models for their behavioral adequacy to real system, and in identification of parameters of the models over the sets of admissible external actions and initial/boundary conditions. The difficulties of solution of these problems increase as the system becomes more complex. For this reason, another important step is structuring of problem domain (or situations), control domain, and simulation scenarios. For these purposes, stratified models, state (or flow) diagrams models, system dynamics models, aggregate models and robust identification can be used. However, the developed model should be subjected to intense analysis and possible changes after its testing for structural controllability, observability, identifiability, and sensitivity. These properties guarantee the model crudity in a given class of variations of the problem conditions and, as a consequence, the reliability and accuracy of system simulation. Besides, the crudity enables one to reduce the model to canonical (more simple) forms which leads to significant simplification of modeling, control synthesis, and analysis of the system.

Thus, when constructing a model of complex dynamical system, three forms of its description arise: (1) conceptual model; (2) formalized model; (3) mathematical model; and (4) computer model.

## 4 Analysis of Complex Systems and their Abstract Description

In system analysis, the problem of structuring of complex systems and processes, their goals, functions, behaviors, etc., is a highly topical problem. Structure is fundamental and, possibly, the most important characteristic of system. Structure can be considered as a set of elements and relations between them, which provides integrity, stability and identity of system under various external and internal changes. One of the main problems of system analysis is building of graphic model describing the existing system of relations and connections between elements. The general problem of analysis is to clarify and establish structural properties of system and its subsystems in a whole, starting from a given description of system elements and relations.

At modern stage of development of systems and control theory, it is more efficient to analyze and model systems not at the level of the systems (control systems) themselves but at the level of the structures of (control) systems. Many definitions of abstract system at the set-theoretic level use ternary description. For example, the notion of abstract system is constructed with use of the following three components: "objects – connections (relations) – properties (attributes)" [A. Hall] or "set of elements – relations – environment" [von Bertalanfi]; summing up various definitions, we can arrive at

$$S =< E, R, L > \qquad (1)$$

where
$E$ is a set of basic elements (subsystems),
$R$ is a set of relations (or connections) between elements,
$L$ is a set of laws and rules that enables constructing different compositions (structures, etc.) with the use of basic elements $E$ and relations $R$.

In some definitions, the set $L$ is substituted by a set of structures $Struct$, which can be considered as a result of action of operator $L$ on the sets $E$ and $R$. Therefore, depending on the chosen basic elements, one can consider several forms of structures of complex system when studying it. This definition can be complemented by other important components, such as parameters, goals, properties, etc. However, it still remains a simplified description of system and it does not satisfy the modern level of knowledge and system requirements. The notion of system has evolved from "elements and relations" to "goals and goal-setting" and then to "observer and language": system is a reflection, on the language of observer (researcher), of objects, relations and their properties in the course of study and cognition.

Let us consider the definition of system, complementing (1) with the following components

$$S =< E, R, Struct, P, W, G, Strat, Rs, S_E > \qquad (2)$$

where
$E$, $R$, $Struct$ are as defined above,
$P$ is a set of parameters of system elements,
$W$ describes integrative properties of system,
$G$ is a set of goals of system functioning,
$Strat$ is a set of strategies of system evolution or development – possible directions, algorithms, mechanisms of self-organization, adaptation,
$Rs$ is a set of resources required for system evolution or development,
$S_E$ is a set of states of environment influencing a system.

Combinations of different components give us some simple models that can be analyzed. For example, the combination $(Struct, E, R)$ reveals a mechanism of formation of different structures from elements $E$ and $R$; the combination $(Struct, P, W)$ is related with the formation of system properties on the basis of parametrization of structure, that is by assigning the elements of structure the certain values of parameters; the combinations $(Struct, E, G)$, $(Struct, P, G)$ and $(Struct, W, G)$ all reveal the influence of both elements of structure and the structure itself and also its parameters and system properties on the formation of goals; the combinations $(Struct, E, W)$ and $(Struct, R, W)$ activate formation of system properties as both elements of structure and structure itself; the combinations $(W, P, G)$, $(W, R, G)$ and $(E, W, G)$ provide a solution for the problem of consistency of system properties and its goals by using structure, parameters, relations, and elements; the combinations $(E, P, Rs, Strat)$ and $(Struct, P, G, Strat)$ determine the strategies of system evolution with elements, resources and changes in values of system parameters, and also interconnect the system evolution with its goals; the combination $(Struct, R, P, S_E)$ reveals the influence of environment on the formation of structures through changes in values of system parameters, appearance of new relations or changes of existing relations in the system. In the same manner, other combinations can be constructed and analyzed.

The model (2) can be further extended by a set of plans $Plans$, which is associated with knowledge-base represented by production rules or semantic networks, and by a set of controls $C$ that influence system evolution and enable purposeful taking a system to a desired state, the point in the system phase space. And finally, we have the following set-theoretic model of complex system

$$S = <E, R, Struct, P, W, G, Plans, Strat, C, Rs, S_E> \quad (3)$$

It is obvious that analysis of complex systems includes analysis of a large number of interrelated combinations. One possible way to reduce the complexity of analysis of the structures is to parametrize the structure of system, elements of system, and relations between them. As an example, the model (2) can be defined in parametrized form as follows

$$S = <Struct(P), W, G, Strat, Rs, S_E> \quad (4)$$

where $Struct(P)$ is the parametrized structure of system

$$Struct(P) = <Struct, R(P), E(P)>$$

where $E(P)$ and $R(P)$ are parametrized characteristics of elements and relations between them. Further, taking into account (4), we get the following model for system (3)

$$S = <Struct(P), W, G, Plans, Strat, C, Rs, S_E> \quad (5)$$

Another way is to classify system components, using certain classification criteria that depend on the purposes of the investigation, and then analyze not the separate components but classes of components. Moreover, complexity is typically reduced by imposing a hierarchical structure on system architecture. As a rule, it is large-scale systems that typically possess hierarchical architecture in order to manage complexity. For systems with hierarchical structure, a connection function between different layers of hierarchy should be defined. One of the main challenges in systems with hierarchical structure is the extraction of a hierarchy of models at various levels of abstraction which are compatible with the functionality and goals of each layer. The connection function determines a rule of how inputs and outputs or state transitions at different levels of hierarchy are connected. This helps to reduce the complexity of analysis of the whole system.

## 5 Mathematical and Computer Simulation of Complex Systems

It is inevitable that when modeling of complex systems one has to deal with aggregate models. For these purposes, the system theoretic notion of an "aggregate" can be used.

Let $T$ be a subset of real numbers (the set of time moments), $X$, $U$, $Y$, $Z$ be sets of any nature. Elements of these sets are: $t \in T$ is a time moment, $x \in X$ is an input signal, $u \in U$ is a control signal (action), $y \in Y$ is an output signal, $z \in Z$ is a state. States, input, control and output signals are considered as functions of time, $z(t)$, $x(t)$, $u(t)$ and $y(t)$.

An aggregate is then defined as
$$S = <T, X, U, Y, Z, H, Q> \quad (4)$$
where $H$ and $Q$ are operators (generally speaking, random operators); $H$ is transition operator, $Q$ is output operator. These operators realize the functions $z(t)$ and $y(t)$. The structure of these operators distinguishes the aggregates among any other systems.

Aggregates are quite general mathematical schemes, special cases of which are Boolean algebras, contact relay networks, finite automata, dynamical systems, and some other mathematical objects.

Based on the object-oriented approach and aggregate model, let us consider the following dynamic model of complex controlled system.

The system is presented as discrete-continuous that reflects the process of state changes in the space of parameters. The parameters vary continuously on an observation time interval $\Delta$. But the system changes its states discretely. The system state is defined as a set of trajectories of parameter changes at time interval $\Delta$. The $q$-dimensional space of parameters is introduced, in which each object at time $t$ is determined by the point
$p(t)=(p^{(1)}(t), p^{(2)}(t),..., p^{(q)}(t))$.
Each parameter is characterized by $v$ variables $p=(\pi_1, \pi_2,..., \pi_v)$. The $i$-th parameter at time $t$ is represented by the point

$p^{(i)}(t)=(\pi_1^{(i)}(t), \pi_2^{(i)}(t),..., \pi_v^{(i)}(t))$, $i=\overline{1,q}$.

Then $p^{(i)}(\Delta_j)=(\pi_1^{(i)}(\Delta_j), \pi_2^{(i)}(\Delta_j),..., \pi_v^{(i)}(\Delta_j))$ defines the trajectory that characterizes the dynamics of $i$-th parameter on time interval $\Delta_j$. Then the state $S_j$ on time interval $\Delta_j$ is defined in space of parameters as
$S_j=[p^{(1)}(\Delta_j), p^{(2)}(\Delta_j),..., p^{(q)}(\Delta_j)]$. Deformation of trajectories at different time intervals $\Delta_j$ and $\Delta_{j+1}$ formalizes the state transition, $S_j \rightarrow S_{j+1}$. Discrete properties of the system are determined by the necessity to divide the state space into subspaces to reflect the observations that characterize the change of states upon transition from one subspace to another.

The model is presented by the components
$D=\{[0,\tilde{T}], <G>, \vartheta, F, L, Cs\}$.
The components are defined as follows.
$[0,\tilde{T}]$ is a finite modeling interval.
$<G>$ is a set of graph-dynamic models describing the behaviors of subsystems of complex system. In each concrete case, depending on the nature of complex system and its properties, $G$ can be represented by different models; these can be agent-based models automata models, (fuzzy) cognitive maps models, state diagrams models, etc. [4], [5], [17], [18], [23-25], [30], [34]. In general case, $G=(S,X,U,Y,M,H,Q)$, where

$S$ is a state space, $S=(S^1, S^2,..., S^n)$, $n$ is the number of state subspaces, $s \in S^i$ has the structure as defined above, $s(t)$ is trajectory of state changes.

$X$ is a set of inputs, $x=[x^1, x^2,..., x^k]$, where $k$ is the number of input channels.

$U$ is a set of control symbols, $u=[u^1, u^2,..., u^r]$.

$Y$ is a set of outputs.

$M$ is a set of output results of monitoring or diagnosing of object, $m=[m^1, m^2,..., m^l]$.

$H$ is a transition operator; it determines the current states on the basis of the system dynamics history; $\{s^j(t)\}=H(s(0),t)$, $j=\overline{1,d}$ is the number of current states; $H=\{H_X, H_U, H_M\}$, where $H_X, H_U, H_M$ are the operators of synthesis of new initial conditions and new behavior depending on the input data/information, control actions, and information about the results of monitoring or diagnosing, respectively.

$Q$ is an output operator, $y(t)=Q(s(0),t)$.

$\vartheta$ defines the system topology and the structural relations between subsystems $G$, for example, hierarchical structure, networked structure, etc.

$F$ is connection function; it determines how state transitions, inputs and outputs of different subsystems are connected with regard to the structure $\vartheta$.

$L=\{DB, KB, A, R\}$ is the information-mathematical model describing the behavior and dynamics of objects at time intervals between the events:

$DB$ - databases,

$KB$ - knowledge-bases,

$A$ - analytical models of object's dynamics at time intervals,

$R$ - rules constructed by experts and based on $DB$ and $KB$; they give rules for re-estimation of current situation in dynamics of parameters of each of the subsystems, and possible formation of new states.

$Cs$ is a model of controlling scenario of system dynamics; $Cs=\{<G>, I_\vartheta, \Phi, T, AE\}$, where

$<G>$ is as defined above,

$I_\vartheta$ is a structural identifier corresponding to $\vartheta$,

$\Phi: I_\vartheta \rightarrow <G>$ is a functional that assigns a structural identifier number to each model $G$,

$T$ is a time diagram for control symbols $U$; it determines the sequential-parallel process of control symbols entering,

$AE$ is a scheme of after-effect of state transitions in different subsystems in accordance with the connection function $F$.

The model of complex system described is quite general. However, it enables one to model various strategies of system development/evolution. The model allows one to iteratively study the behavior of complex system combining both continuous and discrete changes in system dynamics under a number of events associated with input symbols, control actions, monitoring data, etc., and reduces a system dimensionality and simplifies system analysis by specific definition of state.

Because of using of aggregate models, the necessity of development of special computer tools which are adequate to such a description arises. This can be reached by construction of the so-called aggregate simulation systems.

The developed model can be realized in automated computer system of simulation, scenario modeling and control of complex systems. The automated computer system should be based on the comprehensive use of information systems and technologies. The system should have a modular structure that provides sufficient convenience and facility of editing of the separate modules, not influencing the functioning of the others, and adding of new functional capabilities. The information on the problem domain and properties of system under study is contained in specialized databases; knowledge about parameters and processes is contained in knowledge bases; information about the current values of parameters and on the character of state dynamics of object is contained in monitoring databases. The investigations on the models are provided by the combination of the methods of (1) dynamic expert systems, (2) production expert systems, (3) database processing techniques, (4) monitoring/ diagnostic data analysis, (5) scenario control and modeling. The general architecture of computer simulation system (Fig. 1) consists of:

- *User Interface* that provides all functions of computer system associated with exchange of information with user;
- *Parameters Library* that contains the description of parameters for diagnosing and/or monitoring;
- *Knowledge-base* that contains a computer realization of formalized expert knowledge about problem and control domain;
- *Monitoring Databases* that provide the current information about the values of parameters as the result of monitoring or diagnosing;
- *Database of Dynamic Simulation Models* that contains information about the canonical models and templates;
- *Mathematical Modules Library* that contains some typical regularities/laws of system and process dynamics;
- *Scenario Modeling and Control Module* that provides the tools of control scenarios generation, analysis and estimation (Fig. 2); Simulation system of control scenarios that uses the database of control models, the database of system state dynamics simulation models, and the database of control actions provides the tool for formal representation of goals, control problems and system state dynamics, time and resource restrictions. The simulation system of control scenarios enables user to analyze global efficiency of control actions directed on the achievement of goals at different levels of objectives tree. The system allows user to not just compare separate control actions but also to formally synthesize complex control actions (control strategies) for complex system as a whole, and then to further compare them. The functional submodules reproduce the stages of control actions selection, including the stage of synthesis and initiation of problem and control domain and models.
- *Simulation Process Control Module* that serves as a control unit and provides all the stages of the process of system simulation, including expert estimation and comparison of control actions, knowledge-base support, and decision making.

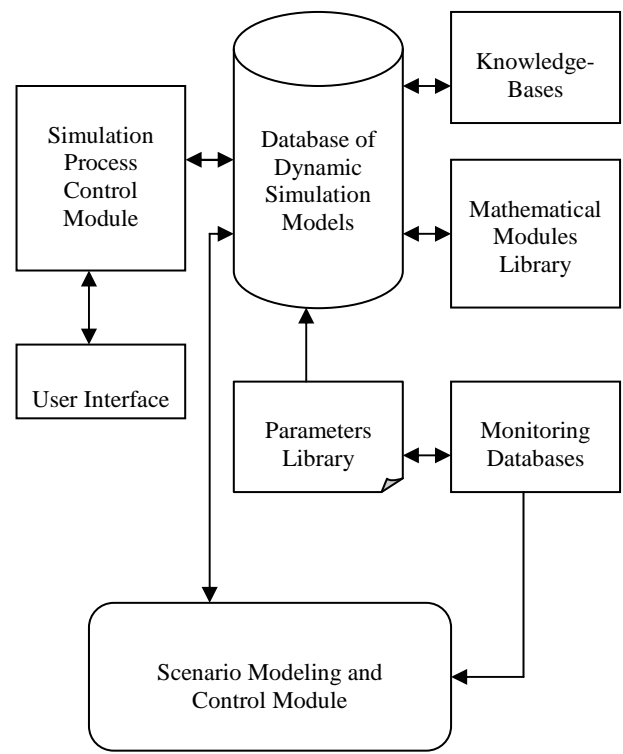

Fig. 1

As it was mentioned earlier, construction of model of complex system consists of three forms of its description: conceptual, formalized, mathematical, and computer ones. Conceptual model reflects informative description of system with use of an informal language. Mathematical description serves as a basis for computer model, which is then transformed to computer simulation system with the help of specialized modeling languages. The main purpose of computer simulation system is to analyze dynamic behavior of system by describing the state changes with time, to predict and compare the consequences of alternative (control) actions or changes in values of system parameters, to evaluate various strategies of system evolution and functioning. Simulation systems provide solution of such problems as evaluation of efficiency of different control principles, comparative analysis of system structures, analysis of the influence of parameters and external conditions on system functioning.

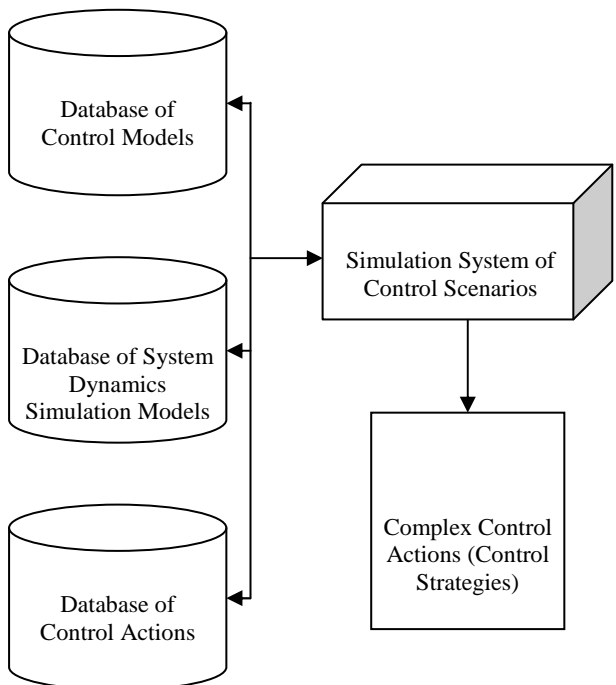

Fig. 2

The basic functions of computer simulation system are the following:
- Identification and registration of the information about the events and the current situation around the complex system. Information and expert knowledge storage, maintaining and management;
- Description of parameters of monitoring and/or diagnosing;
- Description of control actions;
- Description of information-mathematical model;
- Description of the structure of state space;
- Description of the structure of control and problem domains and informational levels of system dynamic models.

The computer system allows user:
- To define global goal of complex system and to represent a decomposition on the subgoals, and to describe relations between them;
- To synthesize dynamic models of complex system;
- To determine efficient control actions in solving control problems and achieving system goals;
- To realize the iterative process of creation and modification of control scenarios;
- To study multi-aspect control of different subsystems and analyze integrative behavior of complex dynamic system as a whole.
- To construct and analyze different control scenarios, study and compare them for efficiency;
- To study behavior of system under different initial conditions and to study state dynamics of system for different groups of parameters.

Simulation system, which as a rule consists of a number of simulation models, should meet the following requirements to:
- *Model completeness*. The models should provide sufficient possibilities for obtaining the necessary characteristics of system with the required accuracy, reliability, and confidence;
- *Model flexibility*. The models should enable one to reproduce various situations upon changing of system parameters;
- *Model structure*. The models should provide the possibility of modification of their separate parts;
- *Information support*. It should provide the information compatibility of models with computer databases.

In the process of development of simulation models two approaches are used: discrete and continuous. Choosing of the approach is determined to much extent by the properties of system and by the character of influence of external environment on the system. For example, Monte-Carlo methods can be considered as a special case of discrete probabilistic simulation models. When using the discrete approach to the development of simulation models, abstract systems (mathematical schemes) of three basic types are normally applied: automata systems, queueing systems, and aggregate systems. In case of continuous approach the system which is modeled is formalized, independently on its nature, in the form of continuous abstract system, between elements of which the flows of one or another nature circulate. The structure of such a system is graphically represented in the form of flow diagrams (schemes).

In general case, simulation model of system can be understood as a system consisting of separate subsystems (elements, components) and connections between them; the functioning (changes of states) and internal change of all the elements of system under the action of connections can have an algorithmic realization, as well as the interaction of system with the environment. Then the simulation of system is reduced to the step-by-step reproduction of the process of functioning of all system elements with regard to their interactions and the influence of the environment. Simulation is more efficient when used at the higher level of hierarchy, where the interaction of a large number of complex objects (processes) is considered on time scales.

Computer simulation system can be developed with the use of universal languages of higher level or with the use of specialized simulation languages. Today, a number of simulation languages and systems are developed, such that SIMULA, GPSS, GASP, SLAM, SIMSCRIPT, POWERSIM, SIMAN, SIMNET, SIMPLE, MODELICA, SIMULINK, and others.

So, the most important components of computer (aggregate) simulation system are the following:
*Simulation model* of complex system (process) along with the software that realizes the computer model;
- *Dialog system* that supports the formation of modeling scenarios;
- *Internal mathematical software* that provides planning and implementation of simulation (computational) experiment;
- *External mathematical software* that provides monitoring and management of the process of simulation experiment;
- *Mathematical software of decision support process* that provides the means of analysis of various control and behavior scenarios and optimal decision making;
- *Standard modules library* containing a set of software modules that implement standard operations;
- *Programming language of higher level* which is used for structural transformation of (aggregate) models, responsible for the formation of input and output information (data) flows.

To make a computer simulation system closer to the end-user in order he/she be capable of conducting experiments independently, the system should be problem-oriented. The development of the corresponding tools of problem-oriented simulation is a vital and topical direction in the field of system modeling technology.

# 6 Conclusion

The basic features and properties of complex systems are considered and analyzed in the context of modeling and simulation. We also discussed in detail the problems of the construction of control and complex system models, and issues of building complex systems as well. We proposed a set theoretic model of complex system. The proposed model is quite general and can be applied for a wide class of systems. The model has been extended in order to be applicable for the case of intelligent complex systems. The proposed dynamic model of complex system can be applied for a wide range of systems, including technical systems, mechanical systems, and it can be easily adapted for socio-economic and strategic planning systems. Much attention has been paid to the development of computer simulation systems and to the analysis of their components. The proposed computer simulation system can be used in applied weakly-formalized control and modeling systems for simulation, analysis, control, and prediction of state dynamics in complex systems. The dynamic models presented can also be used as tools for construction of computer systems for simulation analysis of control strategies and development scenarios of complex objects.